# The orbital and superhump periods of the deeply eclipsing dwarf nova PU UMa

Jeremy Shears, Franz-Josef Hambsch, Colin Littlefield, Ian Miller, Etienne Morelle, Roger Pickard, Jochen Pietz and Richard Sabo

## Abstract

We report unfiltered photometry during superoutbursts of PU UMa in 2009 and 2012. The amplitude was 4.5 magnitudes above mean quiescence and lasted at least 9 to 10 days. Superhumps were present with a peak-to-peak amplitude of up to ~0.3 mag, thereby confirming it to be a member of the SU UMa family of dwarf novae. The mean superhump period during the later part of the 2012 outburst was $P_{sh}$ = 0.08076(40) d. Analysis of the eclipse times of minimum, supplemented with data from other researchers, revealed an orbital period of $P_{orb}$ = 0.077880551(17) d. The superhump period excess was $\varepsilon$ = 0.037(5). During the 2012 outburst, which was the better observed of the two, the FWHM eclipse duration gradually declined from 9.5 to 5 min. The eclipse depth was up to 1.7 magnitudes.

## Introduction

PU UMa was discovered as SDSS J090103.93+480911.1, a cataclysmic variable with high orbital inclination, during the course of the Sloan Digital Sky Survey (1). Deep eclipses were observed in the light curve in a later study by Dillon et al. (2), who also reported the first recorded outburst of the system, confirming it to be a dwarf nova. Their analysis of eclipse minimum times revealed an orbital period of 0.077880505(35) d, although a slightly shorter period of 0.077870588 d could not be ruled out. In quiescence the system had an unfiltered magnitude of 19.5. Observations by the Catalina Real-Time Transient Survey (CRTS) (3) show the system varying between V= 18.7 and 20.2 and with a mean of V=19.5.

We report CCD photometry of outbursts of PU UMa in 2009 and 2012 carried out by a worldwide network of observers using small telescopes.

## Photometry and analysis

Approximately 27 and 95 hours of unfiltered photometry were obtained during the 2009 and 2012 outbursts of PU UMa respectively using the instrumentation shown in Table 1 and according to the observation log in Table 2. Images were dark-subtracted and flat-fielded prior to being measured using differential aperture photometry relative to comparison stars on the BAAVSS chart P090521. We carefully aligned the data from the different observers; adjustments of up to 0.08 mag were made. Heliocentric corrections were applied to all data.

## Outburst light curve

*2009 outburst*

This outburst was detected on 2009 May 16.052 at an unfiltered CCD magnitude of 15.6 (4).The overall light curve of the outburst is shown in Figure 1a. In Figure 2 we plot expanded views of the longer time series photometry runs, where each panel shows one day's data drawn to the same scale. This clearly shows recurrent eclipses superimposed on an underlying modulation, suggestive of superhumps. During the period HJD 2454967 to 9 the system was brightening slightly and this was followed by a plateau phase during which there was a gradual decline over the next 7 days at a mean rate of 0.23 mag/d. Ten days after the outburst was detected the star was at magnitude 18.2, still above the quiescence brightness of mag 19.5 reported by Dillon et al. (2).  At its brightest the star was mag 15.0, representing an outburst amplitude of 4.5 mag.

*2012 outburst*

This outburst was detected on 2012 Feb 1.851 at an unfiltered CCD magnitude of 15.4 (5). The light curve of this outburst is shown in Figure 3a, with expanded views of the photometry shown in Figure 4. We appear to have detected the 2012 outburst at an earlier stage than in 2009. There was a gradual brightening trend over the first 3 days (HJD 2455959 to 62), following which the system faded over the next 6 days at a mean rate of 0.26 mag/d, corresponding to the plateau phase. The outburst was observed over a period of 9 days and at its brightest the star was mag 15.0.

Again eclipses are apparent along with modulations which we interpret as superhumps. The appearance of most of the superhumps was distorted by the presence of eclipses, nevertheless individual superhumps with a peak-to-peak amplitude of 0.3 mag are clearly visible on HJD 2455960. The superhumps declined in amplitude during the outburst and had an amplitude of ~0.18 mag on HJD 2455966. The presence of superhumps is diagnostic that PU UMa is an SU UMa-type dwarf nova. Longer photometry runs were obtained than in the 2009 outburst making the superhumps easier to see and therefore their interpretation more certain.

## Measurement of the orbital period

Times of minimum were measured for nine eclipses observed during the 2009 outburst and a further twenty-three during 2012 using the Kwee and van Woerden method (6) in the *Peranso v2.5* software (7) and are shown in Table 3. In some cases errors were larger due to the eclipse being defined by rather few data points and in other cases the paucity of data prevented a measurement at all. These are supplemented with eclipse times from Dillon et al. (2), which had been used to determine the original value of $P_{orb}$, and four previously unpublished eclipse times from Stuart Littlefair and Christopher Savoury (8).The orbital period was then calculated from an unweighted linear fit to these times of minima as $P_{orb}$ = 0.077880551(17) d. The eclipse time of minimum ephemeris is:

$$HJD_{min} = 2453773.48739(2) + 0.077880551(17) \times E \qquad \text{Equation 1}$$

Our value of $P_{orb}$ is consistent with the longer of the two possible values proposed by Dillon et al. (2). The O-C (Observed – Calculated) residuals of the eclipse minima relative to the ephemeris in Equation 1 are given in Table 3.

**Measurement of the superhump period**

Analysis of the superhumps was complicated by the presence of the eclipses, which as noted before often distorted the shape of the superhumps and prevented accurate measurements of the times of maximum for individual humps. Therefore we used the Lomb-Scargle period analysis algorithm in the *Peranso* software to investigate the superhump periodicity. Before performing the analysis we removed the local trend of the individual data sets. Figure 5a shows the resulting power spectrum of the data from 2012 between HJD 2455959 and 2455962, the interval corresponding to the brightening phase. This has its highest peak at 12.796(37) cycles/d, which we interpret as the orbital signal due to the eclipses, corresponding to $P_{orb}$ = 0.07815(23) d. The error estimates were derived using the Schwarzenberg-Czerny method (9). This value of $P_{orb}$ is close to the one we determined by analysing the eclipse times of minimum, although not coincident. Period analysis, such as Lomb-Scargle analysis, can be affected by other large scale variations in the light curve, hence we prefer to adopt the value of $P_{orb}$ obtained from the time of eclipse minimum analysis. We then pre-whitened the power spectrum with the orbital signal in the frequency domain using *Peranso*. The resulting power spectrum shown in Figure 5b. In this case the strongest signal was at 12.291(49) cycles/d. We interpret this as the superhump signal which corresponds to a superhump period of $P_{sh}$ = 0.08136(33) d. Performing a similar Lomb-Scargle analysis followed by pre-whitening of the data during the slow fade, between HJD 2455962 and 2455965, yielded a superhump signal at 12.383(62) cycles/d and $P_{sh}$ = 0.08076(40) d (power spectrum not shown). Although this might suggest a decrease in $P_{sh}$ during the course of the superoutburst, a phenomenon which is commonly seen in other SU UMa systems (10), a definitive conclusion cannot be drawn given the observed errors in $P_{sh}$.

To check the robustness of our determination of $P_{sh}$ via the pre-whitening method, we applied a second approach. We again took the data between HJD 2455959 and 2455962, but this time we removed the data points corresponding to the eclipses before carrying out a Lomb-Scargle period analysis. The power spectrum (data not shown) had its highest peak at 12.362(67) cycles/d, or 0.08109(41) d. The latter value is similar to the superhump period obtained via the pre-whitening approach.

We also analysed the data from the 2009 outburst (power spectra not shown). Since there were fewer data available, we analysed the combined data from the outburst. Lomb-Scargle analysis resulted in the strongest signal at 12.8571(357) cycles/d, due to the eclipses. Removing this signal via pre-whitening resulted in a spectrum with its strongest signal at 12.369(36) cycles/d. By analogy with the 2012 outburst, again we interpret this signal as due to superhumps, with $P_{sh} = 0.08085(23)$ d. The values of $P_{sh}$ for the 2009 and 2012 outbursts are consistent and we therefore conclude that both events were superoutbursts.

**Analysis of the eclipses**

One of the most interesting aspects of PU UMa is its deep eclipses. We measured the eclipse duration at full width at half minimum (FWHM; Table 3), although the data sampling rate during some of the 2009 eclipses was insufficient to make such measurements. Figure 3c shows that the eclipse duration was greatest in the early stage of the 2012 outburst (10.5 min) and declined as the outburst progressed, with the final eclipses being about one-half the duration (5 min). This is a common feature of eclipses during dwarf nova outbursts and is due to the accretion disc being largest near the start of the outburst and subsequently shrinking from the outside inwards as material drains from the disc as the outburst progresses (11).

There was also a trend of increasing eclipse depth during both the 2009 and 2012 outbursts. The deepest eclipses in 2009 were 2 magnitudes (Figure 1b and Table 3). In 2012 the deepest eclipses were 1.7 magnitudes (Figure 3b and Table 3). A cursory examination of the time series light curves from both outbursts shows that the eclipse depth is also affected by the location of the superhump: in general eclipses are shallower when hump maximum coincides with eclipse.

**Discussion**

Taking our measured orbital period, 0.077880551(17) d and the superhump period measured during the later part of the 2012 superoutburst, $P_{sh} = 0.08076(40)$ d, we calculate the superhump period excess as $\varepsilon = 0.037(5)$. Such a value is consistent with other SU UMa systems of similar orbital period (12).

Patterson *et al.* (13) established an empirical relationship between $\varepsilon$ and q, the secondary to primary mass ratio: $\varepsilon = 0.18*q + 0.29*q^2$. This assumes a white dwarf of ~0.75 solar masses which is typical of SU UMa systems. Our value of $\varepsilon = 0.037$ allows us to estimate q = 0.16.

Dillon et al. (2) noted the apparent similarity between PU UMa and the dwarf nova BG Ari. Both have similar orbital periods ($P_{orb} = 0.082417(28)$ d for BG Ari (2)) and spectra, which suggest that both have spectral type M6 donor stars. Recently, photometry during superoutbursts of BG Ari was presented (14) which confirmed its SU UMa classification and revealed a superhump period excess $\varepsilon = 0.030(7)$. The estimated value of q = 0.14 is similar that for PU UMa, although the outburst amplitude was slightly larger (5.1 mag).

## Conclusions

Our observations of the outbursts of PU UMa in 2009 and 2012 showed that the amplitude was 4.5 magnitudes above mean quiescence and lasted at least 9 to 10 days. Superhumps were present with a peak-to-peak amplitude of up to ~0.3 mag, thereby confirming it to be a member of the SU UMa family of dwarf novae. The mean superhump period during the later part of the 2012 outburst was $P_{sh}$ = 0.08076(40) d. Analysis of the eclipse times of minimum, supplemented with data from other researchers, allowed us to measure the orbital period as $P_{orb}$ = 0.077880551(17) d. The superhump period excess was $\varepsilon$ = 0.037(5), from which we estimated the secondary to primary mass ratio, q = 0.16. During the 2012 outburst, which was better observed than the one in 2009, the FWHM eclipse duration gradually declined from 9.5 to 5 mins and the eclipse depth was up to 1.7 magnitudes.

## Acknowledgements

The authors thank Dr. Stuart Littlefair and Mr. Christopher Savoury of the Department of Physics and Astronomy at the University of Sheffield, UK, for allowing us to use their previously unpublished eclipse timing data from ULTRACAM . JS acknowledges the use of the Bradford Robotic Telescope, on Mt. Teide, Tenerife, operated by the Department of Cybernetics, University of Bradford, UK, with which the 2009 outburst of PU UMa was detected. CL thanks the University of Notre Dame for the use of University's 28 cm SCT. We acknowledge the use of data from the Catalina Real-Time Transient Survey. This research made use of SIMBAD, operated through the Centre de Données Astronomiques (Strasbourg, France), and the NASA/Smithsonian Astrophysics Data System.

We thank our referees, Dr. Chris Lloyd and Dr. Robert Smith, for their helpful comments that have improved the paper.

## References

1. *Szkody P. et al., AJ, 126, 1499-1514 (2003).*

2. *Dillon M. et al., MNRAS, 386, 1568-1578 (2008).*

3. *Drake A.J. et al., ApJ, 696, 870 (2009).*

4. *Shears J., CVnet-outburst group (2009) http://tech.groups.yahoo.com/group/cvnet-outburst/message/3113.*

5. *Shears J., CVnet-outburst group (2012) http://tech.groups.yahoo.com/group/cvnet-outburst/message/4514.*

6. *Kwee K.K. and van Woerden H., Bull. Astron. Inst. Netherlands, 12, 327-330 (1956).*

7. *Vanmunster T., Peranso (2011) http://www.peranso.com/.*

8. *Littlefair S.P. & Savoury C.D., Personal communication (2012).*

9. *Schwarzenberg-Czerny A., MNRAS, 253, 198 (1991).*

10. *Kato T. et al., PASJ, 61, S395–S616 (2009).*

11. *Patterson J. et al., PASP, 112, 1584-1594 (2000).*

12. *Hellier C., Cataclysmic variable stars: How and why they vary, Springer-Verlag (2001) – see Chapter 2.*

13. *Patterson J. et al., PASP, 117, 1204-1222 (2005).*

14. *Shears J. et al, Accepted for publication in JBAA (2011) http://arxiv.org/abs/1109.4133.*

## Addresses

JS: "Pemberton", School Lane, Bunbury, Tarporley, Cheshire, CW6 9NR, UK [bunburyobservatory@hotmail.com]

FJH: Vereniging voor Sterrenkunde, CBA Mol, Belgium, AAVSO, GEOS, BAV [hambsch@telenet.be]

CL: Law School, University of Notre Dame, Notre Dame, IN 46556, USA

IM: Furzehill House, Ilston, Swansea, SA2 7LE, UK [furzehillobservatory@hotmail.com]

EM: Lauwin-Planque Observatory, F-59553 Lauwin-Planque, France [etmor@free.fr]

RP:  3 The Birches, Shobden, Leominster, Herefordshire, HR6 9NG, UK [roger.pickard@sky.com]

JP: Nollenweg 6, 65510 Idstein, Germany [j.pietz@arcor.de]

RS: 2336 Trailcrest Dr., Bozeman, MT 59718, USA [richard@theglobal.net]

| Observer | Telescope | CCD |
|----------|-----------|-----|
| Hambsch | 0.4 m reflector | SBIG STL 11kXM |
| Littlefield | 0.28 m SCT | SBIG ST-8XME |
| Miller | 0.35 m SCT | Starlight Xpress SXVF-H16 |
| Morelle | 0.4 m SCT | SBIG ST-9 |
| Pickard | 0.4 m SCT | Starlight Xpress SXVF-H9 |
| Pietz | 0.20 m SCT | SBIG ST-6B |
| Sabo | 0.43 m reflector | SBIG STL-1001 |
| Shears | 0.28 m SCT | Starlight Xpress SXVF-H9 |

**Table 1: Equipment used**

| Start time (UT) | Start time (JD) | End time (JD) | Duration (h) | Observer |
|---|---|---|---|---|
| **2009 outburst** | | | | |
| May 16 | 2454968.364 | 2454968.430 | 1.6 | Pietz |
| May 17 | 2454968.657 | 2454968.796 | 3.3 | Sabo |
| May 18 | 2454969.688 | 2454969.833 | 3.5 | Sabo |
| May 18 | 2454970.336 | 2454970.433 | 2.3 | Morelle |
| May 19 | 2454970.367 | 2454970.504 | 3.3 | Pietz |
| May 20 | 2454971.336 | 2454971.430 | 2.3 | Morelle |
| May 21 | 2454972.341 | 2454972.412 | 1.7 | Morelle |
| May 22 | 2454974.383 | 2454974.571 | 4.5 | Pietz |
| May 23 | 2454975.426 | 2454975.537 | 2.7 | Miller |
| May 24 | 2454976.443 | 2454976.534 | 2.2 | Miller |
| **2012 outburst** | | | | |
| Feb 1 | 2455959.401 | 2455959.525 | 3.0 | Miller |
| Feb 1 | 2455959.463 | 2455959.569 | 2.5 | Pickard |
| Feb 2 | 2455960.265 | 2455960.707 | 10.6 | Hambsch |
| Feb 2 | 2455960.305 | 2455960.495 | 4.6 | Miller |
| Feb 3 | 2455961.323 | 2455961.390 | 1.6 | Shears |
| Feb 3 | 2455961.335 | 2455961.431 | 2.3 | Pickard |
| Feb 4 | 2455961.743 | 2455962.014 | 6.5 | Sabo |
| Feb 5 | 2455962.686 | 2455962.871 | 4.4 | Littlefield |
| Feb 5 | 2455962.714 | 2455962.987 | 6.6 | Sabo |
| Feb 6 | 2455963.718 | 2455963.993 | 6.6 | Sabo |
| Feb 6 | 2455964.291 | 2455964.672 | 9.1 | Hambsch |
| Feb 8 | 2455965.679 | 2455965.951 | 6.5 | Sabo |
| Feb 8 | 2455966.247 | 2455966.647 | 9.6 | Hambsch |
| Feb 9 | 2455966.539 | 2455966.649 | 2.6 | Littlefield |
| Feb 9 | 2455967.294 | 2455967.669 | 9.0 | Hambsch |
| Feb 10 | 2455968.258 | 2455968.685 | 10.2 | Hambsch |

**Table 2: Log of time-series observations**

| Eclipse number | Eclipse minimum (HJD) | Uncertainty (d) | O-C (s) | Eclipse Duration ( min) | Eclipse depth (mag) | Ref. |
|---|---|---|---|---|---|---|
| 0 | 2453773.48757 | | 16 | | | (2) |
| 1 | 2453773.56532 | | 4 | | | (2) |
| 2 | 2453773.64332 | | 15 | | | (2) |
| 397 | 2453804.40632 | 0.00001 | 30 | | | (8) |
| 411 | 2453805.49671 | 0.00001 | 36 | | | (8) |
| 413 | 2453805.65246 | 0.00001 | 35 | | | (8) |
| 7825 | 2454382.90215 | | -48 | | | (2) |
| 7826 | 2454382.98060 | | 1 | | | (2) |
| 7831 | 2454383.36973 | | -22 | | | (2) |
| 7832 | 2454383.44766 | | -18 | | | (2) |
| 7850 | 2454384.84940 | | -27 | | | (2) |
| 7851 | 2454384.92749 | | -9 | | | (2) |
| 15361 | 2454969.81065 | 0.00032 | 10 | ND | 1.2 | |
| 15368 | 2454970.35561 | 0.00029 | -8 | ND | 1.0 | |
| 15369 | 2454970.43354 | 0.00026 | -3 | ND | 1.0 | |
| 15381 | 2454971.36801 | 0.00045 | -12 | ND | 1.3 | |
| 15394 | 2454972.38023 | 0.00042 | -31 | ND | 1.1 | |
| 15420 | 2454974.40579 | 0.00040 | 26 | ND | 1.6 | |
| 15421 | 2454974.48355 | 0.00034 | 16 | ND | 1.2 | |
| 15434 | 2454975.49538 | 0.00018 | -38 | 3.2 | 2.2 | |
| 15447 | 2454976.50825 | 0.00020 | -1 | 4.0 | 1.9 | |
| 18374 | 2455204.46445 | 0.00004 | -16 | | | (8) |
| 28068 | 2455959.43886 | 0.00021 | 14 | 8.5 | 0.6 | |
| 28069 | 2455959.51665 | 0.00027 | 6 | 9.2 | 0.7 | |
| 28069 | 2455959.51715 | 0.00039 | 49 | 7.1 | 0.7 | |
| 28080 | 2455960.37322 | 0.00024 | -4 | 9.4 | 0.7 | |
| 28081 | 2455960.45152 | 0.00021 | 32 | 6.2 | 0.7 | |
| 28093 | 2455961.38630 | 0.00015 | 51 | 8.3 | 0.8 | |
| 28098 | 2455961.77492 | 0.00054 | -17 | 8.1 | 0.5 | |
| 28099 | 2455961.85263 | 0.00063 | -32 | 8.2 | 0.7 | |
| 28100 | 2455961.93086 | 0.00039 | -1 | 6.3 | 0.7 | |
| 28101 | 2455962.00917 | 0.00048 | 36 | ND | 0.8 | |
| 28110 | 2455962.70984 | 0.00042 | 14 | 6.0 | 1.2 | |
| 28111 | 2455962.78698 | 0.00042 | -50 | 4.3 | 1.6 | |
| 28111 | 2455962.78737 | 0.00024 | -17 | 7.5 | 1.3 | |
| 28112 | 2455962.86553 | 0.00051 | 8 | 4.4 | 1.7 | |
| 28112 | 2455962.86516 | 0.00021 | -24 | 7.8 | 1.3 | |
| 28113 | 2455962.94354 | 0.00042 | 19 | 7.5 | 1.0 | |
| 28124 | 2455963.79997 | 0.00072 | -3 | 7.3 | 0.8 | |
| 28125 | 2455963.87789 | 0.00027 | 0 | 7.9 | 1.1 | |
| 28126 | 2455963.95561 | 0.00042 | -14 | 6.6 | 1.1 | |
| 28149 | 2455965.74734 | 0.00030 | 27 | 5.0 | 0.8 | |
| 28150 | 2455965.82481 | 0.00060 | -9 | 4.8 | 0.9 | |
| 28151 | 2455965.90244 | 0.00078 | -30 | 4.9 | 0.9 | |
| 28160 | 2455966.60371 | 0.00020 | -1 | 4.6 | 1.7 | |

**Table 3: Eclipse minimum times, depth and duration**

Data from the present study, unless referenced otherwise. ND = not determined

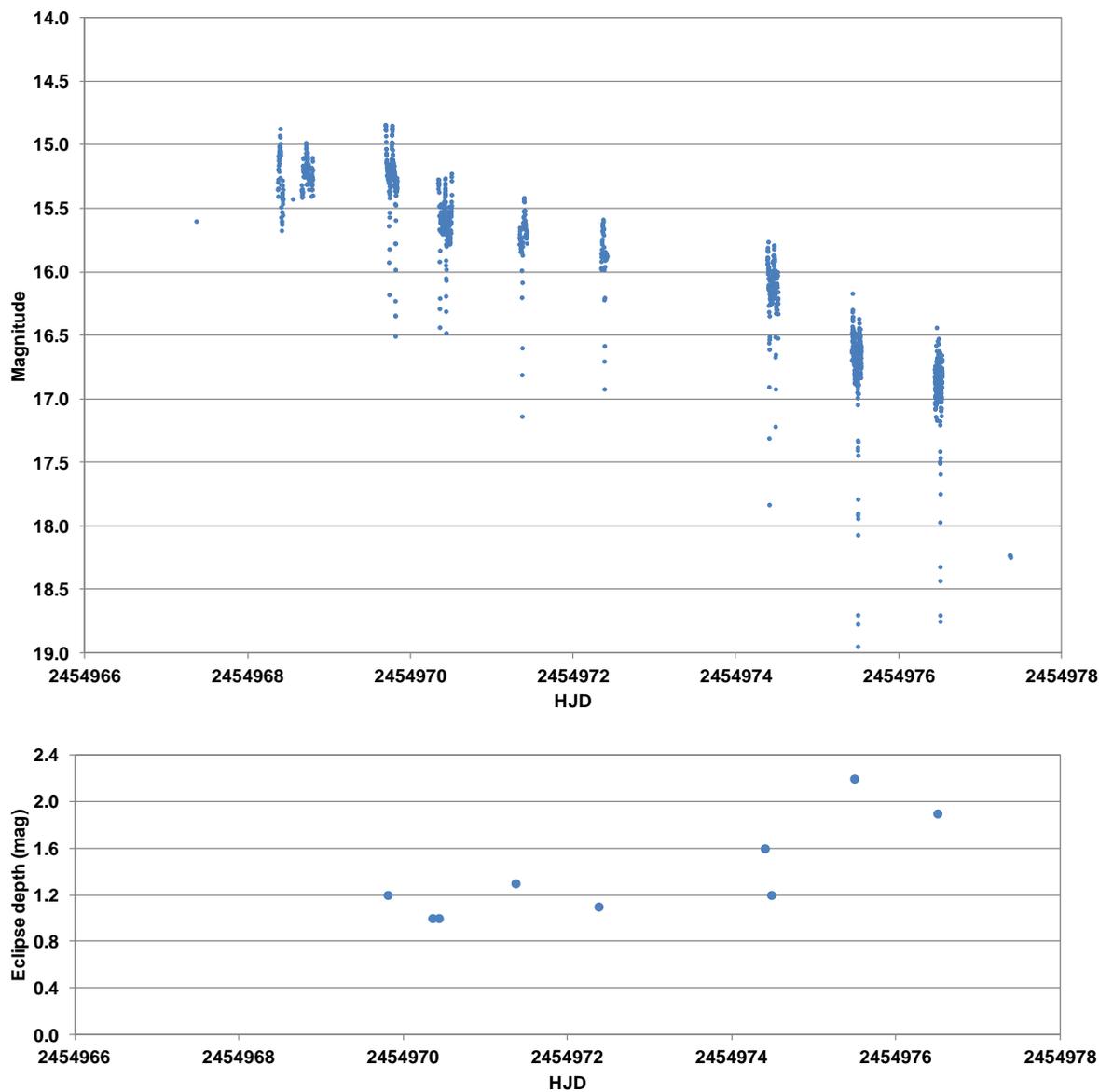

**Figure 1 2009 outburst (a) –top- outburst light curve, (b) – bottom- eclipse depth**

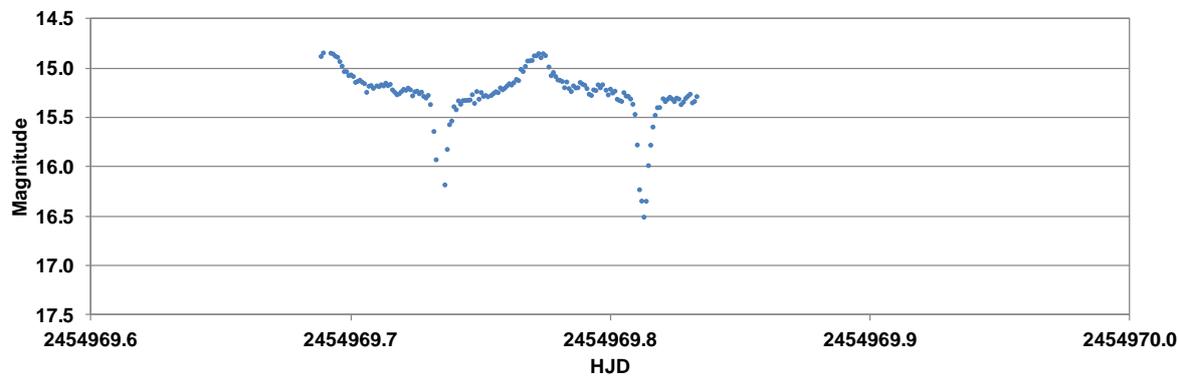

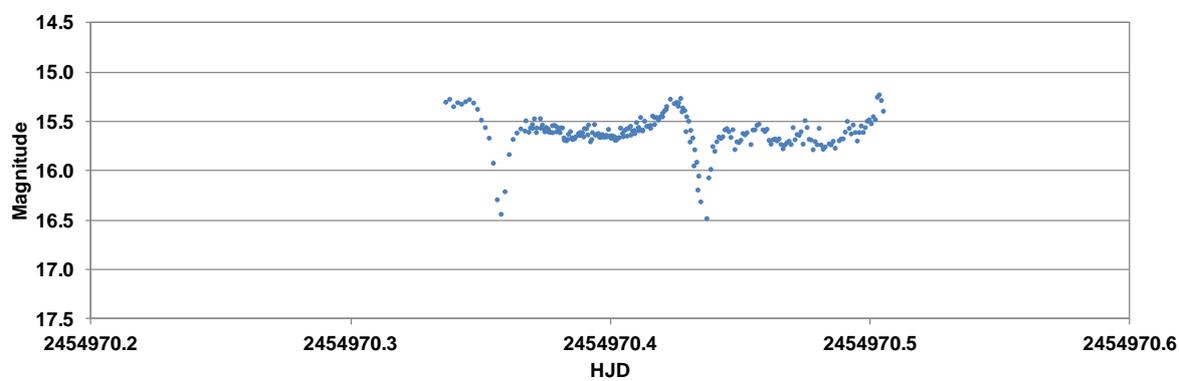

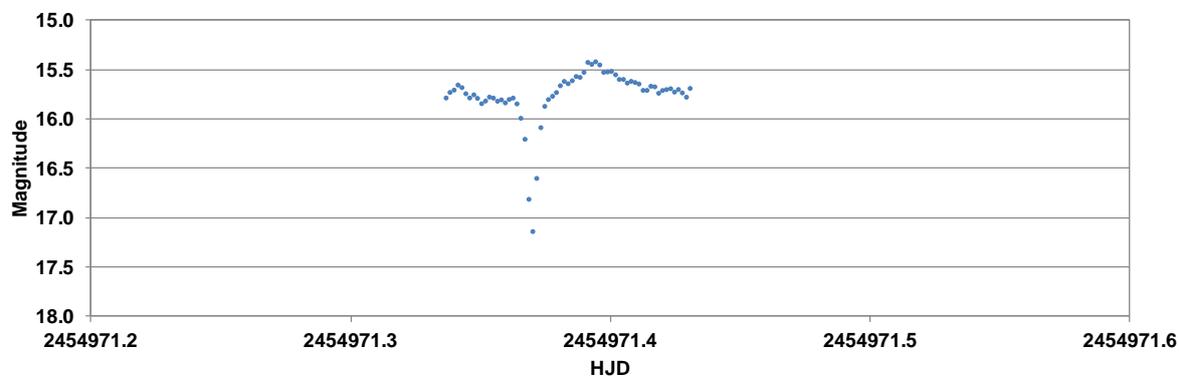

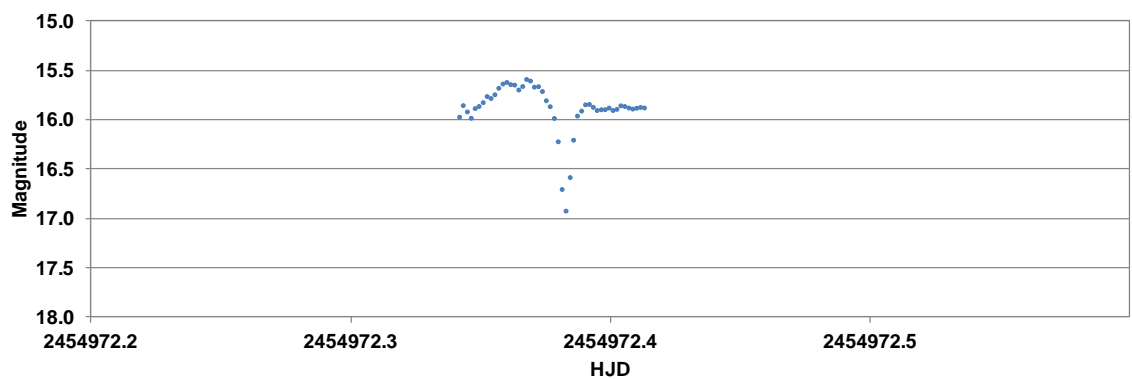

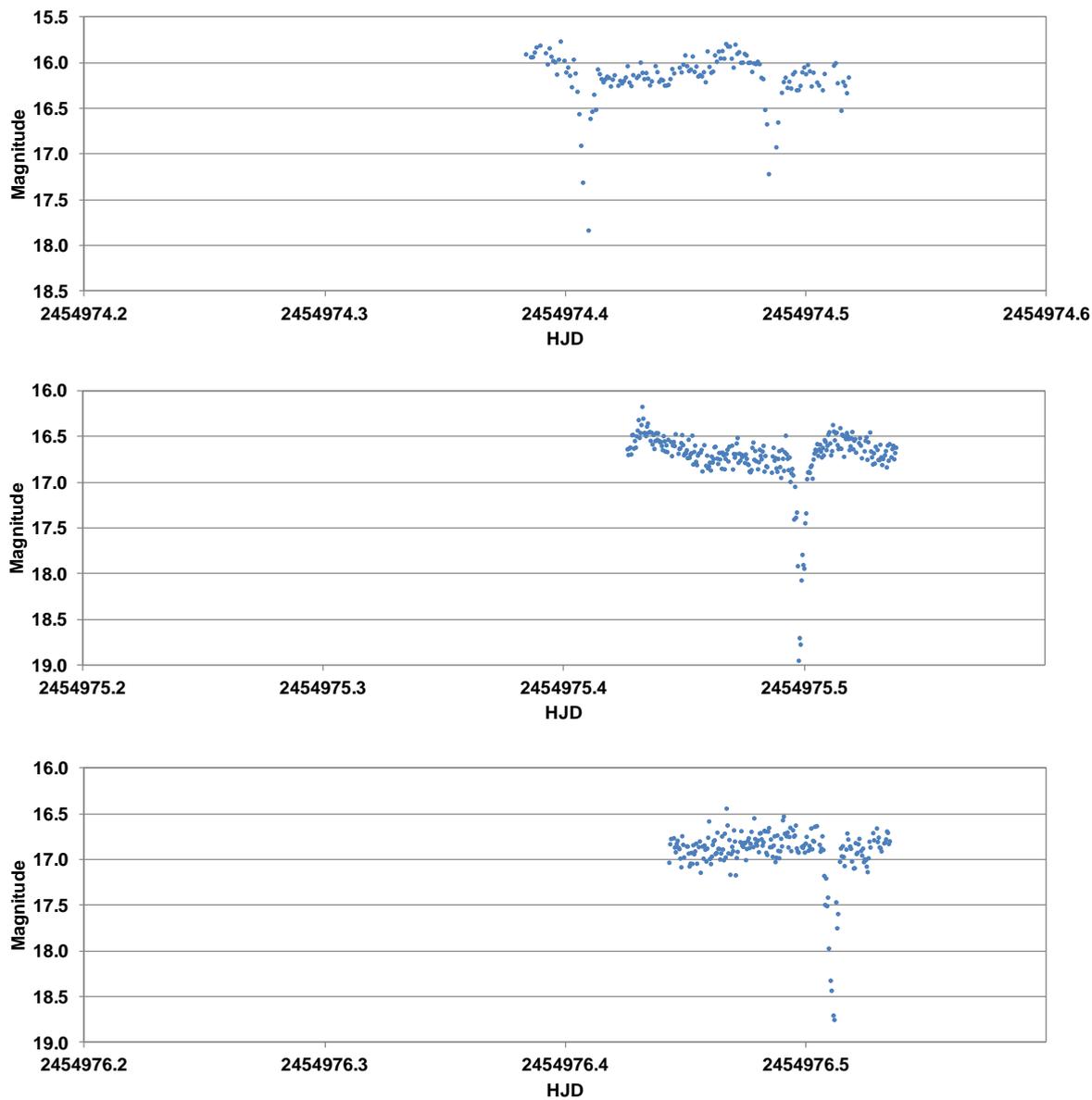

**Figure 2: Expanded views of the time series photometry during the 2009 outburst**

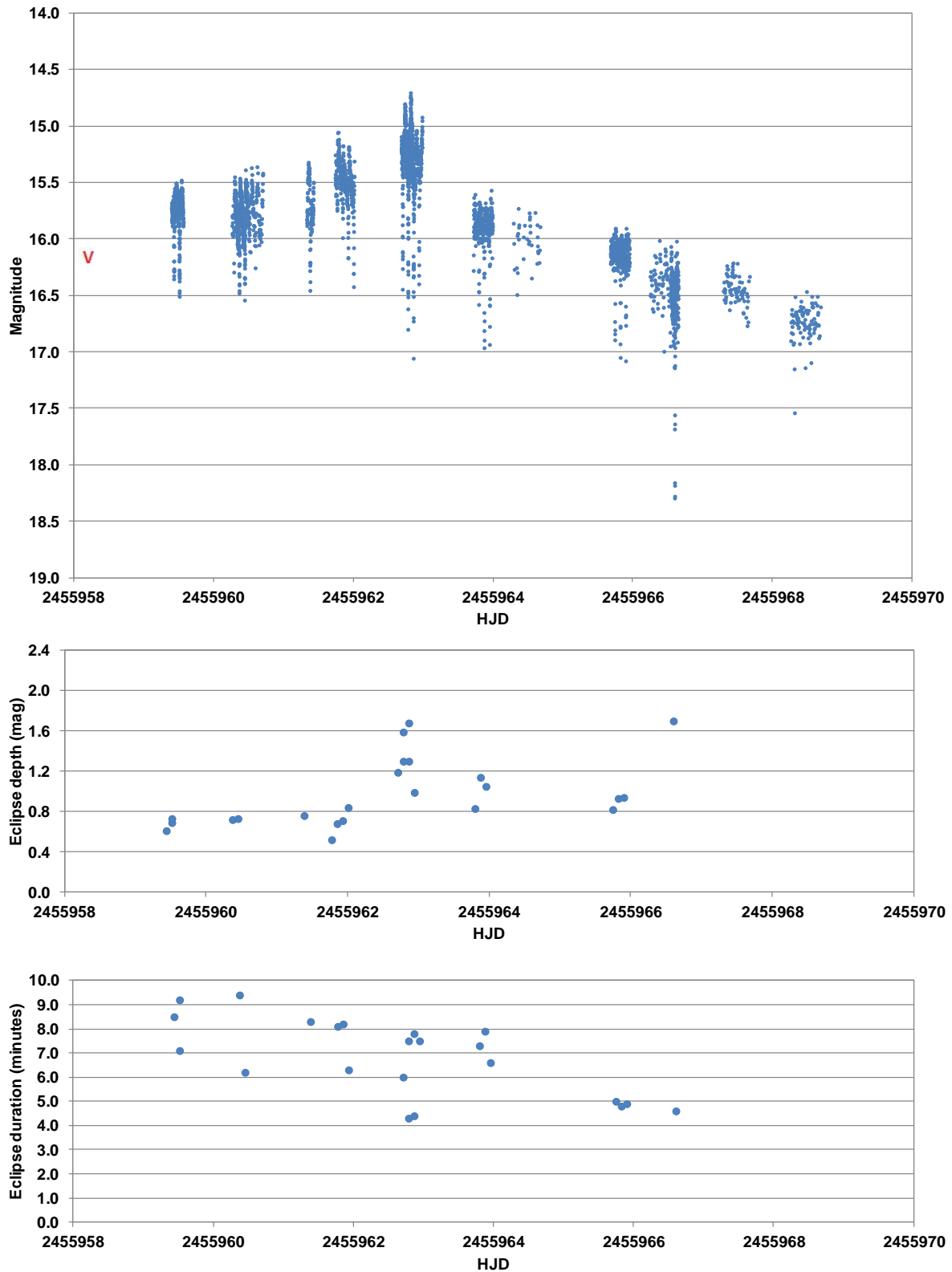

**Figure 3: 2012 outburst (a) – top- outburst light curve, (b) -middle - eclipse depth, (c) – bottom- eclipse duration. The** V **symbol indicates lower magnitude limit**

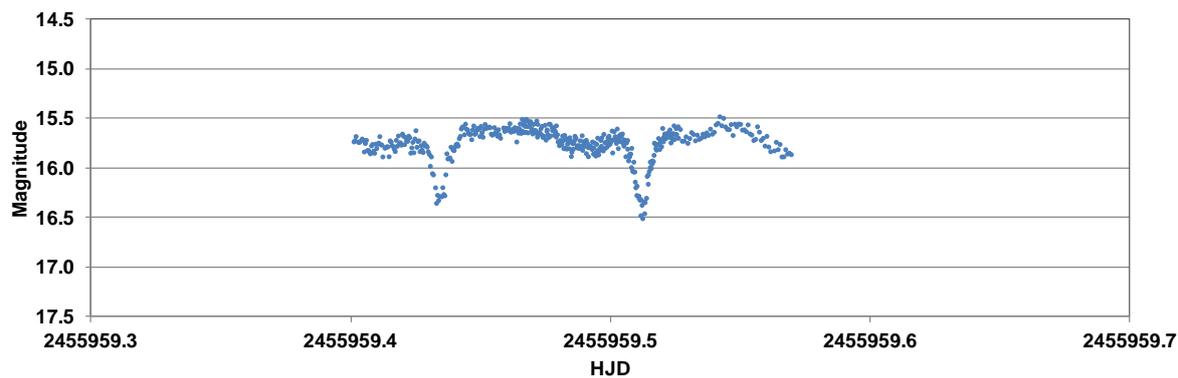

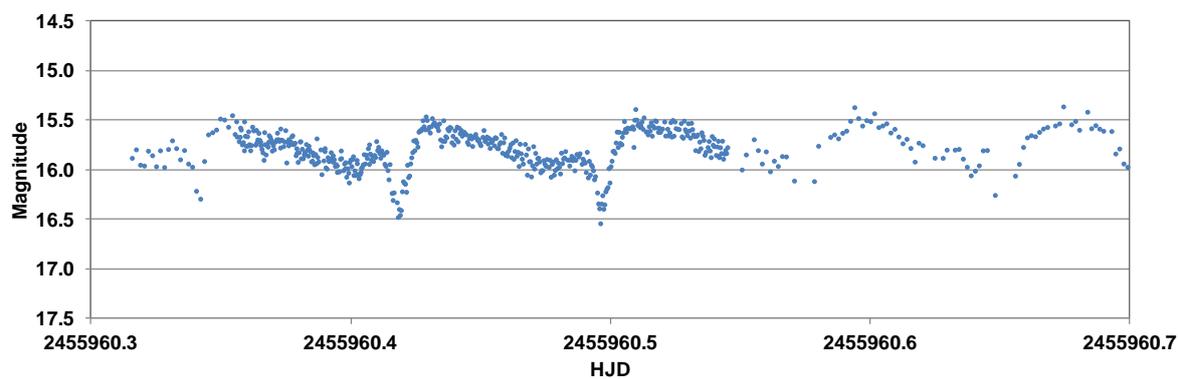

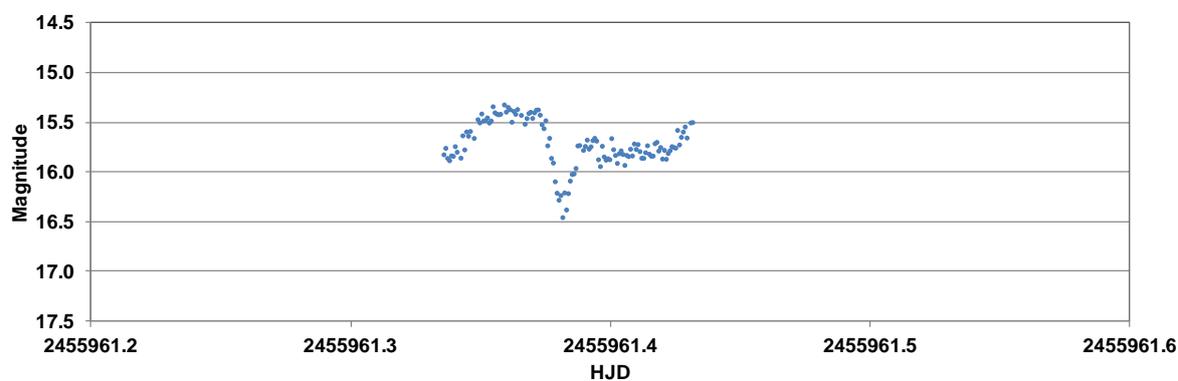

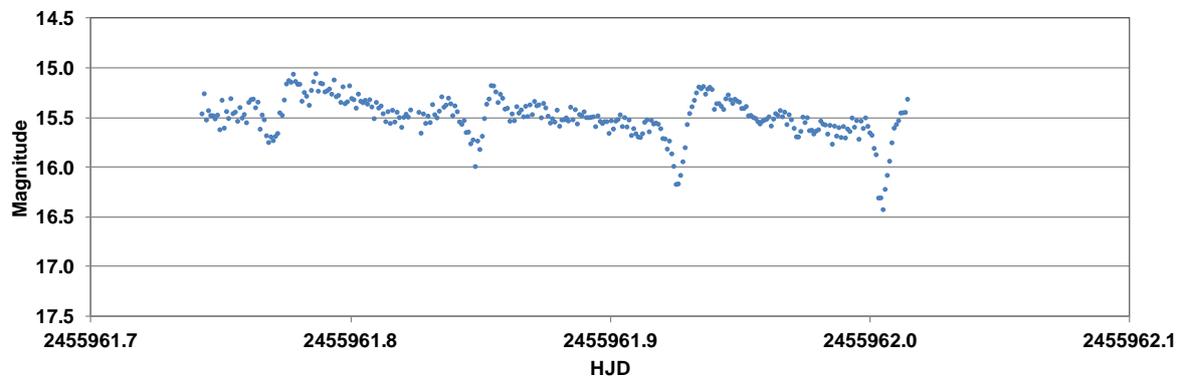

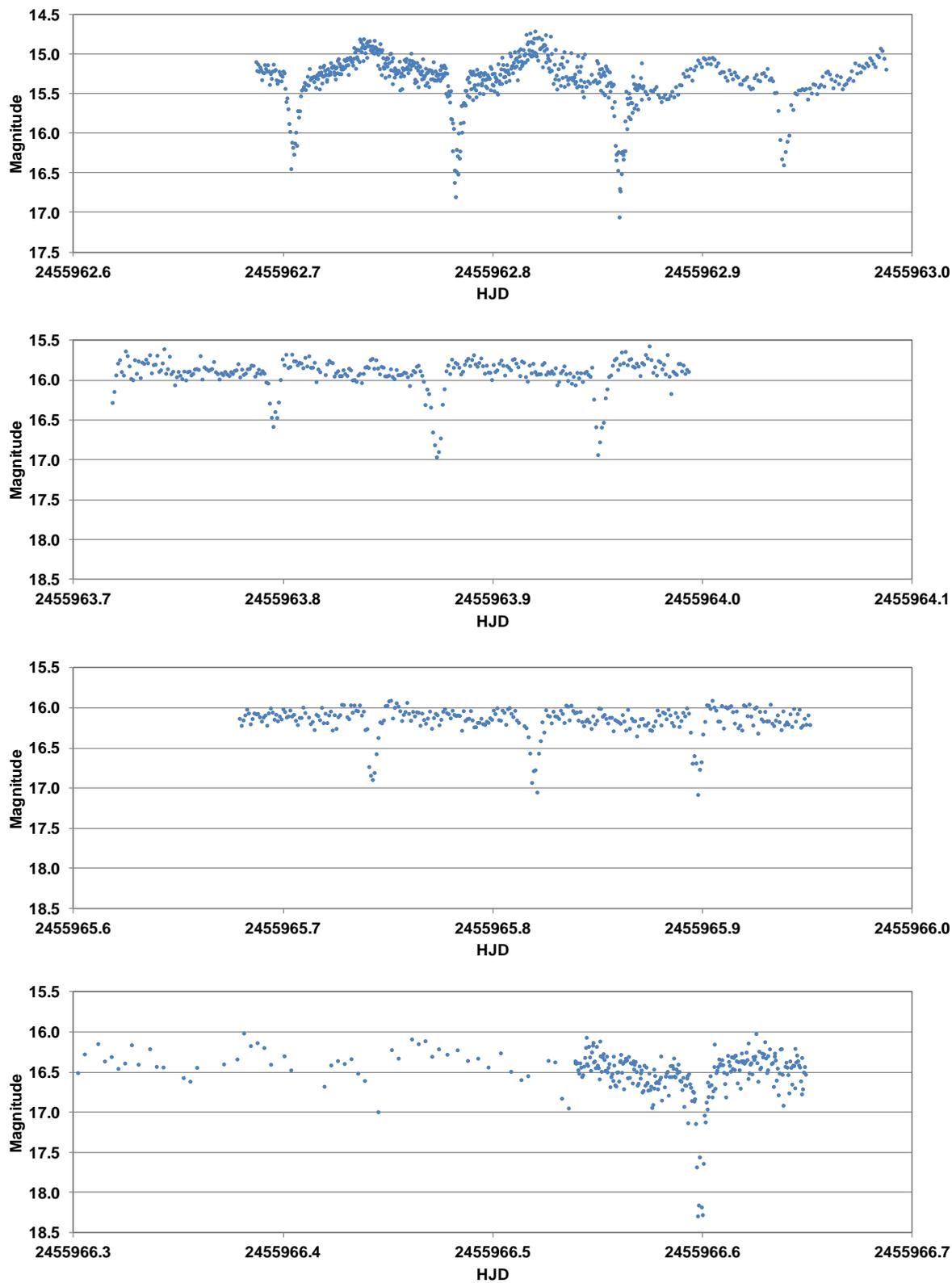

**Figure 4: Expanded views of the time series photometry during the 2012 outburst**

Power

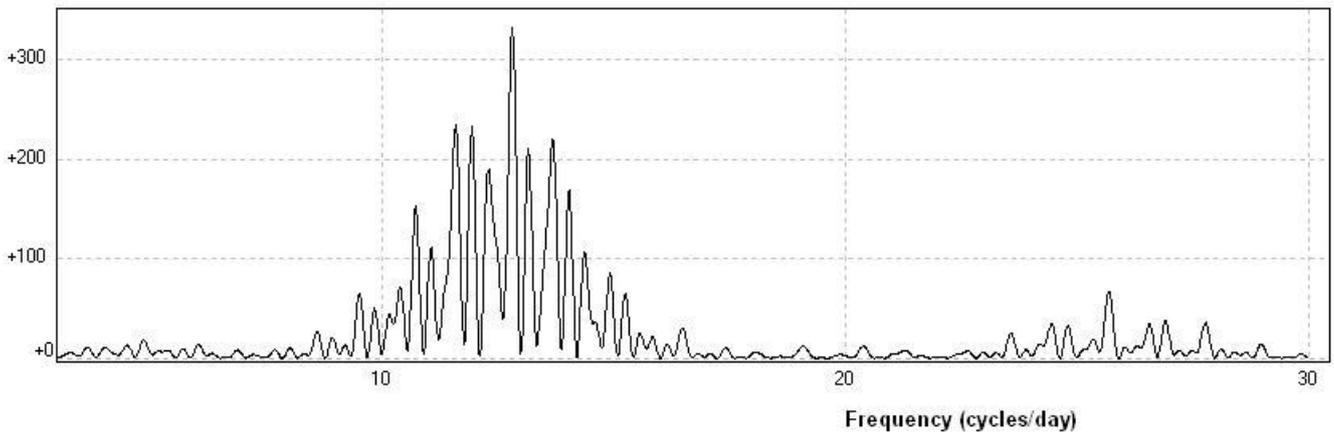

(a)

Power

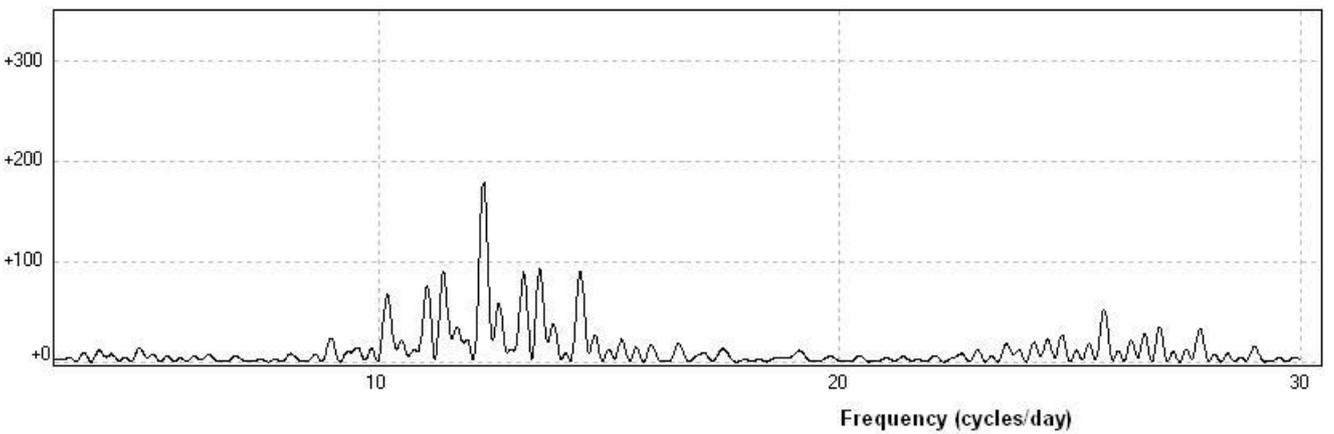

(b)

**Figure 5: (a) Lomb-Scargle power spectrum of the data from the 2012 outburst from HJD 2455959 and 2455962, (b) Lomb-Scargle spectrum after pre-whitening with the orbital signal (12.796 cycles/d)**